\documentclass[twocolumn,amsmath,amssymb,showpacs]{revtex4}
\usepackage{epsfig}
\usepackage{graphicx}

\begin{document}

\title{Debye entropic force and modified Newtonian dynamics }

\author{Xin Li$^{1,3}$}
\email{lixin@itp.ac.cn}
\author{Zhe Chang$^{2,3}$}
\email{zchang@ihep.ac.cn}
\affiliation{${}^1$Institute of Theoretical Physics,
Chinese Academy of Sciences, 100190 Beijing, China\\
${}^2$Institute of High Energy Physics, Chinese Academy
of Sciences, 100049 Beijing, China\\
${}^3$Theoretical Physics Center for Science Facilities, Chinese Academy of Sciences}

\begin{abstract}
Verlinde has suggested that the gravity has an entropic origin, and a gravitational system could be regarded as a thermodynamical system. It is well-known that the equipartition law of energy is invalid at very low temperature. Therefore, entropic force should be modified while the temperature of the holographic screen is very low. It is shown that the modified entropic force is proportional to the square of the acceleration, while the temperature of the holographic screen is much lower than the Debye temperature $T_D$. The modified entropic force returns to the Newton's law of gravitation while the temperature of the holographic screen is much higher than the Debye temperature. The modified entropic force is connected with modified Newtonian dynamics (MOND). The constant $a_0$ involved in MOND is linear in the Debye frequency $\omega_D$, which can be regarded as the largest frequency of the bits in screen. We find that there do have a strong connection between MOND and cosmology in the framework of Verlinde's entropic force, if the holographic screen is taken to be bound of the Universe. The Debye frequency is linear in the Hubble constant $H_0$.
\end{abstract}

\maketitle
\section{Introduction}
Gravity, the most universal force in nature, is still unclear in quantum level. The profound relation between black hole entropy and thermodynamics, which proposed by Hawking {\it et al.}\cite{Hawking}, throws new light to the genius of gravity. This fact leads some physicists investigating gravity from a point of view of thermodynamics. Jacobson\cite{Jacobson} derived Einstein's gravitational field equation from the first law of thermodynamics. The further investigation of Padmanabhan\cite{Padmanabhan1} reinterpreted entropy as the equipartition rule of energy. It also provided a thermodynamic interpretation of gravity. These results support that the gravity has an entropic origin\cite{Padmanabhan2}. Recently, Verlinde\cite{Verlinde} proposed an interesting scenario that interpret gravity as entropic force caused by the changes in the information associated with the positions of material bodies. Verlinde's scenario has two important consequences. One is that, by making use of the the entropic force together with the Unruh temperature\cite{Unruh}, one could derive the second law of Newton. The other is that, by making use of the the entropic force together with the holographic principle\cite{Susskind} and the equipartition law of energy, one could derive Newton's law of gravitation and Komar mass\cite{Komar} in relativistic case.

In the framework of entropic force, various aspects of gravitational physics have been discussed. A succinct list includes: a Friedmann equation derived from entropic force\cite{Shu,Cai1}, the implication for loop quantum gravity\cite{Smolin}, the construction of holographic actions from black hole entropy\cite{Makela,Caravelli}, the holographic dark energy derived from the entropic force\cite{LiM}, the correct entropy/area relation in the relativistic case\cite{Tian}, and the modified entropic force derived from the modified the equipartition law of energy\cite{Gao}. More related works can be seen in \cite{Cai,Easson,Banerjee,Padmanabhan3,Konoplya,Sheykhi,Wei,Zhao}

It should be noticed that Verlinde used the free particle's the equipartition law of energy to derive the Newton's law of gravitation. It is well known that the equipartition law of energy does not hold at very low temperatures, and the equipartition law of energy derived from Debye  model is in good agreement with experimental results for most of the solid objects in very low temperatures.
Verlinde got the Newton's law of gravitation with the assumption that each bit on holographic screen is free of interaction. It should be more general that the bits on holographic screen interact each others. In such case, one could anticipate that the Newton's law of gravitation must be modified. For example, Gao\cite{Gao} used the three dimensional Debye model, which modified the equipartition law of energy, to modify the entropic force. Such modification can interpret the current acceleration of the Universe while without
invoking any kind of dark energy.

Since the equipartition law of energy in Verlinde's scenario\cite{Verlinde} is one dimensional free particle's equipartition law of energy, we use the one dimensional Debye model to modified the entropic force. We find that this modified entropic force is able to derive the famous modified Newtonian dynamics(MOND). MOND was constructed to explaining the flat rotational curves of spiral galaxies.

There are a great variety of observations which show that  the
rotational velocity curves of all spiral galaxies tend to some
constant values\cite{Trimble}. These include the Oort discrepancy in
the disk of the Milky Way\cite{Bahcall}, the velocity dispersions of
dwarf Spheroidal galaxies\cite{Vogt}, and the flat rotation curves
of spiral galaxies\cite{Rubin}. These facts violate sharply the prediction of Newtonian dynamics or Newton's gravity.

The most widely adopted way to resolve these difficulties is the
dark matter hypothesis. It is assumed that all visible stars are
surrounded by massive nonluminous matters. Another approach is the famous MOND\cite{Milgrom}. It assumed that the
Newtonian dynamics does not hold in galactic scale. The particular
form of MOND is given as
 \begin{equation}
 \label{MOND}
 \begin{array}{l}
  m\mu\left(\displaystyle\frac{a}{a_0}\right)\mathbf{a}=\mathbf{F},\\[0.4cm]
 \displaystyle\lim_{x\gg1}\mu(x)=1,~~~\lim_{x\ll1}\mu(x)= x,
 \end{array}
 \end{equation}
where $a_0$ is at the order of $10^{-8}$ cm/s$^2$. At beginning, as
a phenomenological model, MOND explains well the flat rotation
curves with a simple formula and a new parameter. In particular, it
deduces naturally a well-known global scaling relation for spiral
galaxies, the Tully-Fisher relation\cite{TF}. By introducing several
scalar, vector and tensor fields, Bekenstein\cite{Bekenstein}
rewrote the MOND into a covariant formulation. He showed that the
MOND satisfies all four classical tests on Einstein's general
relativity in Solar system.

In this paper, we will derive the MOND in the framework of Verlinde's entropic force.

\section{MOND from modified entropic force}
In verlinde's scenario, the change of a particle's position ($\triangle x$) corresponds the change of entropy ($\triangle S$) associated with the information on the boundary. Motivated by Bekenstein's entropic bound, Verlinde postulated that the change of entropy is proportional to the change of a particle's position,
\begin{equation}
\label{entrophy}
\triangle S=2\pi k_B \frac{mc}{\hbar}\triangle x.
\end{equation}
Such particle will experience an effective force when the holographic screen carries a temperature $T$, together with the first law of thermodynamics, the force equals to
\begin{equation}
\label{entropic F}
F\triangle x=T\triangle S.
\end{equation}
According to the Unruh formula, the temperature in (\ref{entropic F}) corresponds the acceleration
\begin{equation}
\label{Unruh T}
k_BT=\frac{1}{2\pi}\frac{\hbar a}{c}
\end{equation}
where $a$ denotes the acceleration. Since the holographic screen carries temperature, the equipartition law of energy gives the relation between the total energy of the screen and the temperature
\begin{equation}
\label{equipartition}
E=\frac{1}{2}Nk_BT,
\end{equation}
where $N$ denotes the number of degrees of freedom (bits) on the screen.
The number of bits $N$ on the screen is assumed to proportional to the area $A$ of the screen
\begin{equation}
\label{bits}
N=\frac{Ac^3}{G\hbar}.
\end{equation}
Verlinde assumed that the energy of the screen is proportional to the mass that would emerge in the part of space enclosed by the
screen
\begin{equation}
\label{energy mass}
E=Mc^2.
\end{equation}
By making use of the Eq. (\ref{entrophy}), (\ref{entropic F}) and (\ref{Unruh T}), one could obtain the second law of Newton $F=ma$.
By making use of the Eq. (\ref{entrophy}--\ref{energy mass}) together with the relation $A=4\pi R^2$, one could obtain Newton's law of gravitation $F=G\frac{Mm}{R^2}$.

In a statistic system, there are interactions among molecules. Thus the equipartition law of energy for free molecule only valid for the situation that the kinetic energy of molecule is much larger than the effective potential of the interaction between each molecule. Therefore, the equipartition law of energy is invalid at very low temperatures. It is found that Debye model, which modified the equipartition law of energy, is in good agreement with experimental results for most solid objects. Following Verlinde's scenario, we know that the gravity can be explained as an entropic force, it means that the gravity may have a statistical thermodynamics explanation. If so, the gravity should modified while the corresponded statistics changes.

Here, we modified the equipartition law of energy as one dimensional Debye model
\begin{equation}
\label{modified equipartition}
E=\frac{1}{2}Nk_BT\mathfrak{D}(y),
\end{equation}
where the one dimensional Debye function is defined as
\begin{equation}
\mathfrak{D}(y)\equiv\frac{1}{y}\int^y_0\frac{z}{e^z-1}dz.
\end{equation}
$y$ is related to the Debye frequency $\omega_D$, and defined as
\begin{equation}
y\equiv\frac{\hbar\omega_D}{k_BT}=\frac{2\pi c\omega_D}{a}.
\end{equation}
Following Verlinde's method, we obtain the modification of Newton's law of gravitation
\begin{equation}
\label{modified gravity}
\frac{GM}{R^2}=a\mathfrak{D}(\frac{2\pi c\omega_D}{a}).
\end{equation}
There lies two limit case for the modification of Newton's law of gravitation (\ref{modified gravity}).
One is the high temperature limit $y\ll1$, the Debye function $\mathfrak{D}(y)$ reads
\begin{equation}
\mathfrak{D}(y)\approx\frac{1}{y}\int^y_0dy=1.
\end{equation}
Then, the modified equipartition law of energy (\ref{modified equipartition}) returns to (\ref{equipartition}). Therefore, at very high temperature, the Newtonian gravity is recovered.  The other is the low temperature limit $y\gg1$, the Debye function $\mathfrak{D}(y)$ reads
\begin{equation}
\mathfrak{D}(y)\approx\frac{1}{y}\int^\infty_0\frac{z}{e^z-1}dz=\frac{\pi^2}{6y}.
\end{equation}
Then, the modification of Newton's law of gravitation (\ref{modified gravity}) reads
\begin{equation}
\label{modified gravity1}
\frac{GM}{R^2}=\frac{\pi}{12c\omega_D}a^2=\frac{a^2}{a_0},
\end{equation}
where the constant $a_0$ is defined as
\begin{equation}
\label{a0}
a_0=\frac{12c\omega_D}{\pi}.
\end{equation}
Combining the Eq. (\ref{modified gravity}) and (\ref{a0}), we find that
\begin{eqnarray}
\mathfrak{D}(y)=\mathfrak{D}(\frac{\pi^2a_0}{6a})&=&\frac{6}{\pi^2}\frac{a}{a_0}\int_0^{\frac{\pi^2a_0}{6a}}\frac{z}{e^z-1}dz
\end{eqnarray}
Defining the function
\begin{equation}
\mu(x)=\frac{6}{\pi^2}x\int_0^{\frac{\pi^2}{6x}}\frac{z}{e^z-1}dz,
\end{equation}
where $x=\frac{a}{a_0}$, one could find that this function has properties
\begin{equation}
\lim_{x\gg1}\mu(x)=1,~~~\lim_{x\ll1}\mu(x)= x.
\end{equation}
Therefore, the modification of Newton's law of gravitation (\ref{modified gravity}) reads
\begin{equation}
\label{MOND gravity}
\frac{GM}{R^2}=a\mathfrak{D}(\frac{2\pi c\omega_D}{a})=a\mu(x).
\end{equation}
Equation (\ref{MOND gravity}) is just the combination of the MOND relation (\ref{MOND}) and Newton's law of gravitation. Thus, the modified entropic force (\ref{modified gravity}) could account for the MOND. To specific this point, by making use of the formula (\ref{MOND gravity}) we give the modified Poisson equation for the gravitational potential $\phi$
\begin{equation}
\label{Poisson}
\nabla\cdot(\mu(|\nabla\phi|/a_0)\nabla \phi)=4\pi G\rho,
\end{equation}
where $\rho$ is the energy density of matter sources. This equation is just modified Poisson equation in MOND\cite{Milgrom1}.

Milgrom\cite{Milgrom,Milgrom1} observed the relation between the MOND constant $a_0$ and the Hubble constant $H_0$ that
\begin{equation}
\label{MOND relation}
2\pi a_0\approx cH_0.
\end{equation}
In MOND\cite{Milgrom1}, this relation (\ref{MOND relation}) is just a coincidences, and it may point to a strong connection between MOND and cosmology. In Verlinde's scenario, we show that this coincidence is just a consequence of modified entropic force.
By making use of the formula (\ref{a0}) and (\ref{MOND relation}), we find that the Debye frequency $\omega_D$ is proportional to the Hubble constant $H_0$
\begin{equation}
\label{rel H0 Wd}
H_0\approx24\omega_D.
\end{equation}
Since the Debye frequency $\omega_D$ is the largest oscillate frequency of the particle on holographic screen, its wave length $\lambda<c/\omega_D$ should be less than scale of the screen. On the other hand, the cosmological horizon $L_h\sim c/H_0$ has the same scale as the holographic screen, for the screen could be regarded as the bound of the Universe. The relation (\ref{rel H0 Wd}) tells us that
\begin{equation}
\lambda<c/\omega_D=c/24H_0<c/H_0.
\end{equation}
 Thus, the relation (\ref{rel H0 Wd}) is reasonable.

\section{conclusion}
In this paper, we find that a modified entropic force is connected with modified Newtonian dynamics. In Verlinde's scenario, gravity has an entropic origin and gravitational system could be regarded as a thermodynamical system. However, it is well-known that the equipartition law of energy is invalid at very low temperature. Therefore, entropic force should be modified while the temperature of the holographic screen is very low. It is shown that the modified entropic force is proportional to the square of the acceleration, while the temperature of the holographic screen is much lower that the Debye temperature $T_D=\frac{\hbar\omega_D}{k_B}$. And the modified entropic force returns to the Newton's law of gravitation while the temperature of the holographic screen is much larger that the Debye temperature. This thermodynamical approach gives a possible origin of the famous MOND. Also, the constant $a_0$ involved in MOND is linear in the Debye frequency $\omega_D$, which can be regarded as the largest frequency of the bits in screen. We find that there do have a strong connection between MOND and cosmology in the framework of Verlinde's entropic force, if the holographic screen is the bound of the Universe. It is shown that the Debye frequency is linear in the Hubble constant $H_0$. A specific microscopic statistical thermodynamical model of spacetime may give the origin of our modified entropic force model and the microscopic origin for the MOND and the cosmology. We will study it in the future.

\vspace{1cm}
\begin{acknowledgments}
We would like to thank Prof. C. J. Zhu, Y. Tian and X. N. Wu for useful discussions. The
work was supported by the NSF of China under Grant No. 10525522 and
10875129.
\end{acknowledgments}

\end{document}